\documentclass[pra,showpacs,bibnotes]{revtex4}
\usepackage{graphicx}
\begin{document}
\draft
\def\ds{\displaystyle}
\title{Dynamical Casimir Effect for Two Oscillating Mirrors in 3-D}
\author{Cem Yuce$^{*}$, Zalihe Ozcakmakli}
\address{Department of Physics, Anadolu University,
Eskisehir, Turkey.\\ cyuce@anadolu.edu.tr}
\email{cyuce@anadolu.edu.tr}
\begin{abstract}
The generation of photons in a three dimensional rectangular
cavity with two moving boundaries is studied  by using the
Multiple Scale Analysis (MSA). It is shown that number of photons
are enhanced for the cavity whose walls oscillate symmetrically
with respect to the center of the cavity. The non-stationary
Casimir effect is also discussed for the cavity which oscillates
as a whole.
\end{abstract} \maketitle

\section{Introduction}

When two perfectly conducting plates are placed close to each
other, the attractive force appears between the parallel
conductors due to the vacuum fluctuations as predicted long time
ago by Casimir \cite{z1}. The corresponding vacuum energies and
forces were static. Now let us assume that the right boundary
depends on time. In this case, the length of the cavity changes in
time, $\ds{L = L(t)}$. The most evident manifestation of dynamic
behavior is the dependence of the force on time. The modified
force depends on time \cite{bordag}.
\begin{equation}\label{bordag}
F=\frac{\pi \hbar c }{24 L(t)}~.
\end{equation}
More strikingly, when the right plate is in motion, it was
theoretically predicted that photons are generated in the empty
cavity, because of the instability of the vacuum state of the
electromagnetic field in the presence of time-dependent boundary
conditions \cite{z2,z3,z4,z31,ekle0,ekle1,ekle2,ekle3,z9,z11}. A
number of virtual photons from the vacuum are converted into real
photons. This phenomenon is known as dynamical Casimir effect or
motion-induced radiation. However, there has been no experimental
verification for this effect up to now because of the technical
difficulties. There are a few proposed experimental for the
detection of photons \cite{qw1,qw2,deney1}. As it was discussed in
the literature, the best way to observe this effect is to vibrate
one of the wall with one of the resonant field frequencies. A one
dimensional cavity with two perfectly parallel reflecting walls,
one of which is motionless and the other oscillating with a
mechanical frequency equal to a multiple of the fundamental
optical resonance frequency of the static cavity, has been used as
a simple model to study the dynamical Casimir effect
\cite{z3,z4,z9,z11}.
\begin{eqnarray}\label{L1L2gdh}
L_{1}(t) = 0~,~~~~~~ L_{2}(t) ~= L\left(1+\epsilon ~ \sin \Omega
t\right)~ ,
\end{eqnarray}
where the constant $\Omega $ is the external frequency,
$\ds{\epsilon}$ is a small parameter, the constant $\ds{L}$ is the
initial length of the cavity and $\ds{L_{1}(t)}$ and
$\ds{L_{2}(t)}$ are the positions of the right and the left walls
at time $\ds{t}$, respectively. The cavity is motionless initially
and that at some instant one mirror starts to oscillate resonantly
with a tiny
amplitude.\\
Calculating the number of generated photons is a difficult task
since one has to solve the wave equation with the time-dependent
boundary conditions. A lot of technique have been developed to
deal with the problem. For example, averaging over fast
oscillations \cite{z24,z26}, multiple scale analysis \cite{z27},
the rotating wave approximation \cite{z28}, numerical techniques
\cite{z29} are
applied to the dynamical Casimir problem.\\
The case of cavities with two moving mirrors was studied by few
authors
\cite{z16,z17,twomoving1,twomoving2,twomoving3,twomoving4,twomoving5,z5}.
Compared to the situation with a single oscillating mirror, it was
found that radiation is resonantly enhanced when the cavity
oscillates as a whole, with its mechanical length kept constant
and when the cavity oscillates symmetrically with respect to the
cavity center. The radiation emitted by two oscillating walls in
one dimension was studied by Dalvit and Mazzitelli using
renormalization group method \cite{twomoving1}, by Dodonov using
the method known in the theory of parametrically excited systems
\cite{twomoving3} and by Lambrecht et al. using the scattering
approach \cite{z5}.\\
In this paper, we will study two moving boundary problems for a
rectangular cavity resonator in $3-D$. We will use multiple scale
analysis (MSA) which provides us with a solution valid for a
period of time longer than that of the perturbative case, which is
not suitable for this problem since it breaks down after small
time because of the resonance terms. We will investigate the two
cases. In the first configuration, we will consider a rectangular
cavity resonator whose right and left walls in the $x$ direction
move in exactly the same way.
\begin{eqnarray}\label{L1L2}
L_{1}(t) = \epsilon ~L\sin \Omega t~,~~~~~~ L_{2}(t) =
L\left(1+\epsilon ~ \sin \Omega t\right)~ ,
\end{eqnarray}
Initially, the length of the oscillating cavity is given by $\ds{
L_{2}(t)-L_{1}(t)=L}$. As time goes on, the length of the cavity
is not changed. This type of motion corresponds to the cavity oscillating as a whole.\\
Secondly, we will consider a rectangular cavity resonator whose
right and left walls in the $x$ direction move opposite to each
other.
\begin{eqnarray}\label{L1L2cydhj}
L_{1}(t) = -\epsilon ~L\sin \Omega t~,~~~~~~ L_{2}(t) =
L\left(1+\epsilon ~ \sin \Omega t\right)~ ,
\end{eqnarray}
We will show that  the number of generated photons are enhanced
for this system with the parametric resonance case, in which the
frequency of the wall is twice the frequency of some unperturbed
mode, say $\ds{\Omega=2 \omega_k}$.\\
This paper is organized as follows. Section II studies the field
quantization in the case of moving boundaries in one dimension.
Section III reviews and applies the multiple scale analysis.
Section IV studies the dynamical Casimir effect for a three
dimensional cavity. Section V considers the cavity whose walls are
oscillating symmetrically with respect to the center of the
cavity. Finally the last section discusses the enhancement of
generated photon number .

\section{FIELD QUANTIZATION WITH FIXED LENGTH}

Consider a one dimensional cavity formed by two perfect
conductors. The right and the left walls oscillate according to
the formula given by (\ref{L1L2gdh}). The cavity oscillates as a
whole. The field operator in the Heisenberg representation
$\ds{\Phi(x, t)}$ obeys the wave equation $\ds{(c=1)}$
\begin{equation}\label{waveequ}
\frac{\partial^{2}\Phi}{\partial
t^{2}}-\frac{\partial^{2}\Phi}{\partial x^{2}}=0~.
\end{equation}
The boundary conditions are given by
\begin{equation}\label{bc}
\Phi(L_1,t)=\Phi(L_2,t)=0~,
\end{equation}
which describes the moving boundary problem in field theory. The
problem looks simple, since the wave equation is the same as
whether the boundaries are moving or not. However, the moving
boundary conditions render the equation unsolvable by the usual
means. Although the solution of the wave equation is easy to be
found and well known, finding the exact solution of the same
problem endowed with the time-dependent boundary conditions is
very difficult and not known except for some special cases.\\
To solve the problem, we will transform the moving boundary
conditions to the fixed boundary conditions. Let us introduce a
coordinate transformation as
\begin{equation}\label{coortrans}
q(t)=\frac{x-L_{1}(t)}{L}~.
\end{equation}
As a result, the new time-independent boundary conditions for
$\ds{\Phi(q,t)}$ read
\begin{equation}\label{bc2}
\Phi(q=0,t)=\Phi(q=1,t)=0~.
\end{equation}
With this coordinate transformation, the wave equation
(\ref{waveequ}) is changed. Let us find the new form of the wave
equation. Under the coordinate transformation (\ref{coortrans}),
the derivative operators transform as
\begin{eqnarray}\label{s4}
\partial_{t}^{2} & \rightarrow & \partial_{t}^{2}+\frac{\dot{L_{1}}^{2}}{L^{2}}~\partial_{q}^{2}-
2\frac{\dot{L_{1}}}{L}~
\partial_{t}~\partial{q}-\frac{\ddot{L_{1}}}{L}~\partial_{q}~, \nonumber\\
\partial_{x}^{2} & \rightarrow &  \frac{1}{L^{2}}~\partial_{q}^{2}~.
\end{eqnarray}
Here we use the notations $\ds{\partial_{t}\equiv \frac{\partial
}{\partial t}}$, $\ds{\partial_{q}\equiv \frac{\partial }{\partial
q}}$. By using the explicit forms of $L_1(t)$ and $L_2(t)$
(\ref{L1L2gdh}), the transformation of the time derivative
operator up to the first order of $\ds{\epsilon}$ can be
approximated as
\begin{equation}\label{s5}
\partial_{t}^{2} \approx  \partial_{t}^{2}-
\epsilon \left( ~2\Omega ~ \cos \Omega t ~
\partial_{t}\partial{q}~~-~~\Omega^2~  \sin \Omega t ~\partial_{q} ~\right)~.
\end{equation}
Substituting these into the wave equation, we get
\begin{equation}\label{s6}
\frac{1}{L^2 }~ \frac{\partial^2 \Phi}{\partial {q}^{2}} =
\frac{\partial^2 \Phi}{\partial {t}^{2}}- \epsilon\left(~ \Omega ~
\cos \Omega t~ \frac{\partial^2 \Phi}{\partial {q}
\partial {t}}-\Omega^2 ~ \sin \Omega t ~\frac{\partial
\Phi}{\partial {q}} ~\right)~.
\end{equation}
In the right hand side of the equation,  the term in the
parenthesis is due to the time dependent boundary conditions. Now,
we don't have to deal with the time dependent boundary conditions.
However, we are left with a new equation.\\
Let us now solve the above equation subject to the fixed boundary
conditions (\ref{bc2}). The field operator can be expanded as
\begin{equation}\label{expansion}
\Phi(q, t)=\sum_{k}~\left(~b_{k}\Psi_{k}(q,
t)+b_{k}^{\dag}\Psi_{k}^{\star}(q, t)~\right)~,
\end{equation}
where $\ds{b_{k}^{\dag}}$and $\ds{b_{k}}$ are the creation and the
annihilation operators, respectively and  $\ds{\Psi_{k}(q, t)}$ is
the corresponding mode function. We will follow the approach given
in \cite{z20,z21,z22} to find the explicit form of functions
$\ds{\Psi_{k}(x, t)}$ . For an arbitrary moment of time, the mode
function satisfying the boundary conditions is expanded as
\begin{equation}
\Psi_{k}(q,t>0)=\sum_{n} a_{n}^{k}(t)~\sin(n\pi q)~,
\end{equation}
Let us substitute it into the equation (\ref{s6}) in order to find
$\ds{a_{n}^{k}(t)}$. Then, multiply the resulting equation with
$~\ds{\sin(m\pi q)}$ and integrate over $q$ from $0$ to $1$. If we
use the orthogonality relations, we get an infinite set of coupled
differential equations for $\ds{a_{n}^{k}}$ after some algebra.
\begin{equation}\label{denklem00}
\ddot{a}_m^{k}+\omega_m^{2}~ a_m^{k}=\epsilon\left(4\Omega \cos
\Omega t \sum_{n\neq m}g_{nm}~\dot{a}_n^{k}-2\Omega^{2} \sin
\Omega t \sum_{n\neq m}g_{nm}~a_n^{k} \right)
\end{equation}
where~ $\ds{ \omega_m=m\pi/L}$ and the antisymmetric coefficient
is given by $\ds{
g_{nm}=\frac{mn(1-(-1)^{m+n} )}{m^2-n^2}}$ for $\ds{m\neq n}$.\\
In the next section, we will solve the above equation. To do this,
we prefer to use multiple scale analysis method.

\section{Multiple Scale Analysis (MSA)}

Conventional weak-coupling perturbation theory suffers from
problems that arise from resonant terms in the perturbation
series. The effects of the resonance could be insignificant on
short time scales but become important on long time scales.
Perturbation methods generally break down after small time
whenever there is a resonance that lead to what are called secular
terms. In the equation (\ref{denklem00}), this happens for those
particular values of external frequency $\Omega$ such that there
is a resonant coupling with the eigenfrequencies of the static
cavity. To avoid such problems, we will use Multiple-scale
analysis (MSA), a powerful and sophisticated perturbative method
valid for longer times. Multiple-scale perturbation theory
provides a good
description of our system.\\
The trick is to introduce a new variable $\ds{\tau=\epsilon t}$.
This variable is called the slow time because it does not become
significant in the small time. The functional dependence of
$\ds{a_{m}^{k}}$ on $t$ and $\epsilon$ is not disjoint because it
depends on the combination of $\epsilon t$ as well as on the
individual $t$ and $\epsilon$. The time variables $t$ and
$\ds{\tau}$ are treated independently in MSA. Thus, in place of
$\ds{a_{m}^{k}(t)}$, we write $\ds{a_{m}^{k}(t,\epsilon t)}$. Let
us expand $\ds{a_{m}^{k}}$ in the form of a power series in
$\epsilon$
\begin{equation}\label{msa1}
a_{m}^{k}(t)=a_m^{k(0)}(t, \tau)+\epsilon~ a_m^{k(1)}(t,
\tau)+\epsilon^2~a_m^{k(2)}(t, \tau)+...~.
\end{equation}
To this end, we change the independent variable in the original
equation from $t$ to $\tau$. Using the chain rule, we have
$\ds{\frac{d}{dt}\rightarrow\frac{\partial}{\partial t}+\epsilon
\frac{\partial}{\partial
\tau}}$. \\
Up to the first order of $\epsilon$, the derivatives with respect
to the time scale $\ds{t}$ are given by
\begin{eqnarray}\label{msa2}
\dot{a}_m^{k} &=& \partial_t~
a_m^{k(0)}+\epsilon~\left(\partial_{\tau}a_m^{k(0)}+\partial_{t}a_m^{k(1)}\right)\nonumber\\
\ddot{a}_m^{k} &=&
\partial_{t}^2~a_m^{k(0)}+\epsilon~\left(2\partial_{\tau
}\partial_{ t}~a_m^{k(0)}+\partial_{t}^2~a_m^{k(1)}\right)~,
\end{eqnarray}
where dot denotes time derivation with respect to $t$ as usual.
Let us substitute these into the equation (\ref{denklem00}). Then,
we see that our original ordinary differential equation is
replaced by a partial differential equation. It may appear that
the problem has been complicated. But, as will be seen below,
there are many advantages of this method. To zeroth order in
$\epsilon$, we get a well-known equation in physics.
\begin{equation}
\ddot{a}_m^{k(0)}+\omega_{m}^2 ~a_m^{k(0)}=0~.
\end{equation}
To first order in $\epsilon$, we obtain the following equation.
\begin{equation}\label{uhgfx588586}
\ddot{a}_m^{k(1)} + \omega_{m}^2 a_{m}^{k(1)}=-2\partial_{
t}\partial_{\tau }~a_{m}^{k(0)}+4\Omega \cos(\Omega t) \sum_{n\neq
m}g_{nm}~ \dot{a}_{n}^{k(0)}-2\Omega^{2} \sin(\Omega t)\sum_{n\neq
m}g_{nm}~a_n^{k(0)}
\end{equation}
The solution of the former one can be found easily
\begin{equation}\label{msacozum1}
a_{m}^{k(0)}(t,\tau)=A_m^{k} (\tau)~e^{-i\omega_{m}t}+B_m^{k}
(\tau) ~e^{i\omega_{m}t}~.
\end{equation}
Note that $A_m^{k} (\tau)$ and $B_m^{k} (\tau)$ are not constants
but functions of the slow time scales $\tau$. The initial
conditions are given by
\begin{eqnarray}\label{i1}
A_{m}^{k}(\tau=0)&=&\frac{1}{\sqrt{2\omega_{k}}}\delta_{m, k}\nonumber\\
B_{m}^{k}(\tau=0)&=&0
\end{eqnarray}
Let us now solve the equation (\ref{uhgfx588586}). We look for the
oscillations of the cavity that could enhance the number of
generated photons by means of resonance effects for some specific
external frequencies. To enhance the number of photons, let us now
assume the resonance condition, $\ds{\Omega=p~\pi/L }$, where
$\ds{p=1,2,...}$. It is well known that, in the resonance
conditions, the number of generated photons grows very
much in time.\\
Let us substitute the zeroth order solution (\ref{msacozum1}) into
the right hand side of the equation (\ref{uhgfx588586}) and then
use the following relations: $\ds{2i\sin \Omega t=(e^{i\Omega
t}-e^{-i\Omega t})}$,  $\ds{2\cos \Omega t=(e^{i\Omega
t}+e^{-i\Omega t})}$. It can be seen that the right hand side
contains terms that produce secular terms. For a uniform
expansion, these secular terms must vanish. In other words, any
term with $\ds {~e^{\pm i\omega_{m}t}~}$ on the right-hand side
must vanish. If not, these terms would be in resonance with the
left-hand side term and secularities would appear. After imposing
the requirement that no term ~$\ds {e^{- i\omega_{m}t}~}$ appear,
we get
\begin{equation}\label{onc1}
\partial_{\tau}A_{m}^{k}~+G_{p+m,m}^{-}~A^{k}_{p+m}~-G^{+}_{m-p,m}~A^{k}_{m-p}
-G^{-}_{p-m,m}~B^{k}_{p-m}=0
\end{equation}
where $\ds{G_{i,j} }$ is defined as
\begin{equation}\label{Gmn}
{G^{\mp}_{i,j}} = \frac{\Omega\mp 2\omega_{i}}{2\omega_{j}}~
\Omega  ~g_{ij} ~.
\end{equation}
In the similar way, the fact that no secularities should arise
from the term with $\ds{ e^{ i\omega_{m} t}  }$ leads to
\begin{equation}\label{onc2}
-\partial_{\tau}~B_{m}^{k}~-G^{-}_{p+m,m}~B^{k}_{p+m}~+~G^{+}_{m-p,m}~B^{k}_{m-p}
+~G^{-}_{p-m,m}~A^{k}_{p-m}=0
\end{equation}
To this end, let us give the formula for the number of generated
photons \cite{z30,z31}
\begin{equation}\label{number1}
\langle N_n \rangle=\sum_{k}2\omega_{n}|B_{n}^{k}|^{2}.
\end{equation}
We will now analyze the solutions of the equations
(\ref{onc1},\ref{onc2}) for a given $p$.

\subsection{Analysis of Solution}

As a special case, let us study the above equations when
$\ds{m=p}$. In this case, the equations (\ref{onc1},\ref{onc2})
are reduced to the following simple ones.
\begin{equation}\label{reduced1}
\partial_{\tau}A_{p}^{k}~+G_{2p,p}^{-}~A_{2p}^{k}=0
\end{equation}
\begin{equation}\label{reduced2}
\partial_{\tau}~B_{p}^{k}~+G^{-}_{2p,p}~B^{k}_{2p}=0
\end{equation}
To find the solution, we should also find the corresponding
differential equations for $\ds{A_{2p}, B_{2p}}$. Substituting
$\ds{m=2p}$ in (\ref{onc1},\ref{onc2}) gives
\begin{equation}\label{reduced11}
\partial_{\tau}A_{2p}^{k}~+G_{3p,2p}^{-}~A_{3p}^{k}
~-G_{p,2p}^{+}~A_{p}^{k}=0
\end{equation}
\begin{equation}\label{reduced22}
\partial_{\tau}~B_{2p}^{k}~+G^{-}_{3p,p}~B^{k}_{3p}~-G_{p,2p}^{+}~B_{p}^{k}=0
\end{equation}
Note that since $p-m\rightarrow p-2p<0$, the term with
$B^{k}_{p-m}$ vanishes. As can be seen easily, we need the
equations $A_{3p}^{k}$ and $B_{3p}^{k}$ to solve these
differential equations. In fact, there is no upper cutoff.
However, these equations are not coupled equations. $A_{m}^{k}$
and $B_{m}^{k}$ are not coupled to each other. So, we say that no
photons with $\ds{\omega_p,\omega_{2p},...}$ are generated for a
given external frequency $\ds{\Omega=\pi p/L}$. But this doesn't
mean that photons are not generated. For example, photons with
$\ds{\omega_{p+1}}$ are produced. To calculate the number of
photons in this mode, we should solve (\ref{onc1},\ref{onc2}) when
$m=p+1$. As was seen above, we are left with infinitely many
equations since there is no cutoff. This is due to the fact that
the spectrum of a one-dimensional cavity is equidistant. Intermode
coupling produces resonant creation in the other modes.
So, the spectrum does not have an upper frequency cutoff. So, multiple scale analysis does not estimate the number of generated photons in one dimension.\\
To sum up, photons are produced resonantly in all modes except
$m=p,2p,...$. In the next section, we will study the dynamical
Casimir effect for a three dimensional cavity whose spectrum is
not equidistant.

\section{Dynamical Casimir Effect in 3-D}

So far, we have restricted ourselves to the one dimensional case.
We will now study the dynamical Casimir effect for the three
dimensional geometries. We will show that multiple scale analysis
works very well to calculate the number of generated photons.\\
Let us firstly define our problem. Consider a rectangular cavity
resonator with perfectly conducting walls. Initially the three of
the them are placed at $x=0, y=0$ and $z=0$ while the other three
walls are at $x=L_x, y=L_y$ and $z=L_z$. At time $t=0$, the cavity
starts oscillating in the x-direction as a whole. The positions of
the six walls at any time are given by
\begin{equation}\label{L1L23D}
L_{1x}(t) = \epsilon ~L_x\sin \Omega t~,~~ L_{2x}(t) =
L_x\left(1+\epsilon~ \sin \Omega
t~\right)~,~~L_{1y}(t)=0~,~~L_{2y}(t)=L_y~,~~L_{1z}(t)=0~,~~L_{2z}(t)=L_z~.
\end{equation}
Here the constant $\Omega $ is the frequency of the oscillation
and $\ds{\epsilon}$ is a small parameter. The volume of the cavity
resonator is constant in time and given
by $V=L_x L_y L_z$.\\
As in the one dimensional case, we will study with the coordinates
transformed from the fixed ones to the moving ones,
$(x,y,z)\rightarrow(q,y,z)$. The relation between $q$ and $x$ is
given by $\ds{q=(x-L_{1_{x}})/L_{x}}$.\\
We will firstly study the dynamical Casimir effect for the scalar
field and then for the vector field.

\subsection{Scalar Field in 3-D}

In this section, we will apply MSA to the scalar field in three
dimensions. The cavity resonator oscillates in the x-direction
(\ref{L1L23D}). So, the scalar field operator subject to the
following boundary conditions
\begin{equation}\label{scalarbc}
\Phi(L_{1x},y,z,t)=\Phi(L_{2x},y,z,t)=\Phi(x,0,z,t)=\Phi(x,L_y,z,t)=\Phi(x,y,0,t)=\Phi(x,y,L_z,t)=0~,
\end{equation}
which describes the moving boundary problem in three dimensions.
The field operator in the Heisenberg  representation
$\ds{\Phi(x,y,z,t)}$ obeys the wave equation $\ds{(c=1)}$ in three
dimensions
\begin{equation}\label{waveequ3d}
\nabla^2 \Phi=\frac{\partial^{2}\Phi}{\partial t^{2}}~.
\end{equation}
We can rewrite this equation in the moving coordinate systems,
$\ds{(q,y,z)}$. The transformation of the time derivative operator
is given in (\ref{s5}). Hence, the wave equation is transformed to
\begin{equation}\label{scalars6}
\frac{\partial^2 \Phi}{\partial {y}^{2}}+\frac{\partial^2
\Phi}{\partial {z}^{2}}+\frac{1}{L^2 }~ \frac{\partial^2
\Phi}{\partial {q}^{2}} = \frac{\partial^2 \Phi}{\partial
{t}^{2}}- \epsilon\left( 2\Omega~ \cos (\Omega t)~
\frac{\partial^2 \Phi}{\partial {q} \partial {t}}-\Omega^2 \sin
(\Omega t) ~\frac{\partial \Phi}{\partial {q}} \right)~.
\end{equation}
In one dimension, the expansion of the solution of the wave
equation was given by the formula (\ref{expansion}). In three
dimension, it has obvious generalization. The scalar field
operator can be expanded as
\begin{equation}\label{scalarexpansion}
\Phi(q,y,z,
t)=\sum_{k_x,k_y,k_z}~b_{k_{x}k_{y}k_{z}}~\Psi_{k_{x}k_{y}k_{z}}(q,y,z,t)+H.C.~.
\end{equation}
where $\ds{b_{k_{x}k_{y}k_{z}}}$ is the annihilation operator and
$\ds{\Psi_{k_{x}k_{y}k_{z}}(q,y,z,t)}$ is the corresponding mode
function. Let us assume that, for an arbitrary moment of time, the
explicit form of function $\ds{\Psi_{k_{x}k_{y}k_{z}}(q,y,z,t)}$
is expanded as
\begin{equation}
\Psi_{k_{x}k_{y}k_{z}}(q,y,z,t>0)=\sum_{n_x,n_y,n_z}
a_{n_{x}n_{y}n_{z}}^{k_{x}k_{y}k_{z}}(t)~\sin(n_{x}\pi
q)~\sin(\frac{n_{y}\pi}{L_y}y)~\sin(\frac{n_{z}\pi}{L_z} z)~.
\end{equation}
Substituting these into the transformed wave equation
(\ref{scalars6}) and using the orthogonality relations, we get an
infinite set of coupled differential equations for
$\ds{a_{n_{x}n_{y}n_{z}}^{k_{x}k_{y}k_{z}}(t)}$ after some
algebra.
\begin{equation}\label{denklem00gf}
\ddot{a}_{m_{x}n_{y}n_{z}}^{k_{x}k_{y}k_{z}}+\omega_{m_{x}n_{y}n_{z}}^{2}
a_{m_{x}n_{y}n_{z}}^{k_{x}k_{y}k_{z}}=\epsilon\left(4\Omega \cos
(\Omega t) \sum_{n_x\neq
m_x}g_{n_{x}m_{x}}~\dot{a}_{n_{x}n_{y}n_{z}}^{k_{x}k_{y}k_{z}}-2\Omega^{2}
\sin (\Omega t) \sum_{n_x\neq
m_x}g_{n_{x}m_{x}}~a_{n_{x}n_{y}n_{z}}^{k_{x}k_{y}k_{z}} \right)
\end{equation}
where~ $\ds{
\omega_{m_{x}n_{y}n_{z}}^{2}=\pi^2(~{m_{x}^2/L_x^2+n_{y}^2/L_y^2+n_{z}^2/L_z^2}~)}$
and $\ds{ g_{n_xm_x}=\frac{m_x~n_x(1-(-1)^{m_x+n_x}
)}{m_x^2-n_x^2}}$ for $\ds{m_x\neq n_x}$. In fact, this equation
is the three dimensional generalization of (\ref{denklem00}).\\
We will apply the multiple scale analysis to solve this equation.
Fortunately, no need to go into the detail. The generalization of
the equation (\ref{uhgfx588586}) to three dimensions is
straightforward.
\begin{eqnarray}\label{uhdn886}
\ddot{a}_{m_{x}n_{y}n_{z}}^{{k_{x}k_{y}k_{z}(1)}} +
\omega_{m_{x}n_{y}n_{z}}^2~
a_{m_{x}n_{y}n_{z}}^{{k_{x}k_{y}k_{z}(1)}}=&&-2\partial_{\tau
}\partial_{ t}~a_{m_{x}n_{y}n_{z}}^{{k_{x}k_{y}k_{z}(0)}}+ 4\Omega
\cos (\Omega t) \sum_{n_x\neq
m_x}g_{n_{x}m_{x}}~\dot{a}_{n_{x}n_{y}n_{z}}^{k_{x}k_{y}k_{z}
(0)}\nonumber\\
&&- 2\Omega^{2} \sin (\Omega t) \sum_{n_x\neq
m_x}g_{n_{x}m_{x}}~a_{n_{x}n_{y}n_{z}}^{k_{x}k_{y}k_{z}(0)}~,
\end{eqnarray}
where
\begin{equation}\label{msaghcks}
a_{m_{x}n_{y}n_{z}}^{{k_{x}k_{y}k_{z}(0)}}=A_{m_{x}n_{y}n_{z}}^{{k_{x}k_{y}k_{z}}}
(\tau)~\exp{(-i~\omega_{m_{x}n_{y}n_{z}}~t)}+B_{m_{x}n_{y}n_{z}}^{{k_{x}k_{y}k_{z}}}
(\tau) ~\exp{(i~\omega_{m_{x}n_{y}n_{z}}~t)}~,
\end{equation}
$\ds{A_{m_{x}n_{y}n_{z}}^{{k_{x}k_{y}k_{z}}} (\tau)}$ and
$\ds{B_{m_{x}n_{y}n_{z}}^{{k_{x}k_{y}k_{z}}} (\tau)}$ depend on
the slow time parameter $\ds{\tau}$. The initial conditions are
given by
\begin{eqnarray}\label{sgdhjk}
A_{{n_{x}n_{y}n_{z}}}^{{k_{x}k_{y}k_{z}}}(\tau=0)&=&\frac{1}{\sqrt{2\omega_{{k_{x}k_{y}k_{z}}}}}~\delta_{k_x,
n_x} ~\delta_{k_y,n_y}~\delta_{k_z,n_z}~;\nonumber\\
B_{{n_{x}n_{y}n_{z}}}^{{k_{x}k_{y}k_{z}}}(\tau=0)&=&0~.
\end{eqnarray}
We will derive an equation like (\ref{onc1},\ref{onc2}).
Substituting (\ref{msaghcks}) into (\ref{uhdn886}) and eliminating
the secular terms from the equation, we get
\begin{eqnarray}\label{yhfbcnjk}
\partial_{\tau}A_{m_{x}n_{y}n_{z}}^{{k_{x}k_{y}k_{z}}}+G^{-}_{(n_{x}^{\prime} n_{y}n_{z}),(m_{x}n_{y}n_{z})}~A_{n_{x}^{\prime}n_{y}n_{z}}^{{k_{x}k_{y}k_{z}}}-
G^{+}_{(n_{x}^{\prime \prime}n_{y}n_{z}),(m_{x}n_{y}n_{z})}~A_{n_{x}^{\prime \prime}n_{y}n_{z}}^{{k_{x}k_{y}k_{z}}}-G^{-}_{(n_{x}^{\prime \prime \prime}n_{y}n_{z}),(m_{x}n_{y}n_{z})}~B_{n_{x}^{\prime \prime \prime}n_{y}n_{z}}^{{k_{x}k_{y}k_{z}}}&=&0~~~~~~~~\\\nonumber\\
\label{yhfbcnjk2}
-\partial_{\tau}B_{m_{x}n_{y}n_{z}}^{{k_{x}k_{y}k_{z}}}
-G^{-}_{(n_{x}^{\prime}n_{y}n_{z}),(m_{x}n_{y}n_{z})}~B_{n_{x}^{\prime}n_{y}n_{z}}^{{k_{x}k_{y}k_{z}}}+
G^{+}_{(n_{x}^{\prime
\prime}n_{y}n_{z}),(m_{x}n_{y}n_{z})}~B_{n_{x}^{\prime
\prime}n_{y}n_{z}}^{{k_{x}k_{y}k_{z}}}+G^{-}_{(n_{x}^{\prime
\prime \prime}n_{y}n_{z}),(m_{x}n_{y}n_{z})}~A_{n_{x}^{\prime
\prime \prime}n_{y}n_{z}}^{{k_{x}k_{y}k_{z}}}&=&0~~~~~~~~
\end{eqnarray}
where the three dimensional generalization of the definitions
$\ds{G^{\mp}_{ij}}$ (\ref{Gmn}) is given by
\begin{equation}\label{scalarGmn}
{G^{\mp}_{(r_x,n_{y},n_{z}),(m_{x},n_{y},n_{z})}} =~
\frac{\Omega\mp
2\omega_{r_{x}n_{y}n_{z}}}{2\omega_{m_{x}n_{y}n_{z}}}~\Omega~g_{r_xm_{x}}
~.
\end{equation}
It should be noted that $\ds{n_{x}^{\prime }, n_{x}^{\prime
\prime}, n_{x}^{\prime \prime \prime}}$ are positive integers and
they satisfy
\begin{eqnarray}\label{sghddvbcnjk}
\omega_{n_{x}^{\prime }n_{y}n_{z}}&=&~~\Omega+\omega_{m_{x}n_{y}n_{z}}~,\nonumber\\
\omega_{n_{x}^{\prime \prime}n_{y}n_{z}}&=&-\Omega+\omega_{m_{x}n_{y}n_{z}}~,\nonumber\\
\omega_{n_{x}^{\prime \prime \prime
}n_{y}n_{z}}&=&~~\Omega-\omega_{m_{x}n_{y}n_{z}}~.
\end{eqnarray}
In one dimensional case, $\ds{A_m, A_{m\mp p}, A_{m\mp 2p},...}$
and $\ds{B_m, B_{m\mp p}, B_{m\mp 2p},...}$ are strongly coupled
to each other. However, in three dimensions, only a few modes are
coupled to each other. This is because there are only a few
positive integers $\ds{n_{x}^{\prime }, n_{x}^{\prime \prime},
n_{x}^{\prime \prime \prime}}$ satisfied by the equations
(\ref{sghddvbcnjk}). So, it is possible to solve the equations
(\ref{yhfbcnjk},\ref{yhfbcnjk2}) exactly since only
a few modes are coupled.\\
In what follows, we will give some specific examples.

\subsubsection{Examples}

For the simplicity, assume that the cavity is cubic,
$\ds{L_x=L_y=L_z}$. We have a freedom to choose the coupled modes.
For example, assume that inter-mode coupling occurs between
$\ds{(n_{x}^{\prime \prime \prime},n_{y},n_{z})}$ and
$\ds{(m_{x},n_{y},n_{z})}$. So we can determine $\ds{\Omega}$ from
(\ref{sghddvbcnjk}). As an example, we are interested in the
following two modes: $\ds{(1,1,1)}$ and $\ds{(2,1,1)}$. Choose
$\ds{\Omega=(\sqrt{3}+\sqrt{6})\pi/L_x}$. Let us solve the
equations (\ref{yhfbcnjk},\ref{yhfbcnjk2}) for these modes. Hence,
\begin{eqnarray}\label{scalaramc1}
\partial_{\tau}A_{211}^{{k_{x}k_{y}k_{z}}}-G^{-}_{(1,1,1),(2,1,1)}~B_{111}^{{k_{x}k_{y}k_{z}}}&=&0~;\nonumber\\
-\partial_{\tau}B_{211}^{{k_{x}k_{y}k_{z}}}+G^{-}_{(1,1,1),(2,1,1)}~A_{111}^{{k_{x}k_{y}k_{z}}}&=&0~.
\end{eqnarray}
To solve the above differential equations, we also need the
equations for $\ds{\partial_{\tau}A_{1,1,1}}$ and
$\ds{\partial_{\tau}B_{1,1,1}}$. Using again
(\ref{yhfbcnjk},\ref{yhfbcnjk2}), we get
\begin{eqnarray}\label{scalaramc2}
\partial_{\tau}A_{111}^{{k_{x}k_{y}k_{z}}}-G^{-}_{(2,1,1),(1,1,1)}~B_{211}^{{k_{x}k_{y}k_{z}}}&=&0~;\nonumber\\
-\partial_{\tau}B_{111}^{{k_{x}k_{y}k_{z}}}+G^{-}_{(2,1,1),(1,1,1)}~A_{211}^{{k_{x}k_{y}k_{z}}}&=&0~.
\end{eqnarray}
The solution of the equations (\ref{scalaramc1},\ref{scalaramc2})
with the boundary conditions (\ref{sgdhjk}) can readily be found.
\begin{eqnarray}
\left(%
\begin{array}{c}
  A_{111}^{{111}} \\
  \\
  A_{211}^{{211}} \\
\end{array}%
\right)=
\left(%
\begin{array}{c}
  \frac{1}{\sqrt{2\omega_{111}}}\cosh(\lambda\tau)\\
  \\
  \frac{1}{\sqrt{2\omega_{211}}}\cosh(\lambda\tau)\\
\end{array}%
\right)~;~~~~~~\left(%
\begin{array}{c}
  B_{111}^{211} \\\\
  B_{211}^{111} \\
\end{array}%
\right)=
\left(%
\begin{array}{c}
  \frac{G^{-}_{(2,1,1),(1,1,1)}}{\lambda
  \sqrt{2\omega_{211}}}\sinh(\lambda\tau)\\\\
  \frac{G^{-}_{(1,1,1),(2,1,1)}}{\lambda\sqrt{2\omega_{111}}} \sinh(\lambda\tau)\\
\end{array}%
\right)~,
\end{eqnarray}
where the constant $\lambda$ is defined as
$\ds{\lambda^{2}=G^{-}_{(2,1,1),(1,1,1)}~G^{-}_{(1,1,1),(2,1,1)}}$.
With the help of number operator (\ref{number1})
\begin{equation}
\langle N_{n_xn_yn_z}\rangle=\sum_{k_x} \sum_{k_y}\sum_{k_z}
 2~\omega_{n_xn_yn_z}~|B_{n_xn_yn_z}^{{(k_{x}k_{y}k_{z})}}|^{2}
\end{equation}
we find the number of generated photons for each mode
\begin{eqnarray}
\langle N_{1,1,1}\rangle &=&  ~\sinh^{2}(\lambda~\tau)\nonumber\\
\langle N_{2,1,1}\rangle &=& ~\sinh^{2}(\lambda~\tau).
\end{eqnarray}

\subsection{Vector Field}

Here, we will study the dynamical Casimir effect for the vector
field in three dimensions. Maxwell's equations describe
electromagnetic waves as having two components, the electric
field, $E(x, y, z)$, and the magnetic field, $H(x, y, z)$. Modes
in a cavity resonator are said to be transverse. It is convenient
to classify the fields as transverse magnetic $(TM)$ or transverse
electric $(TE)$ according to whether E or H was transverse to the
direction of oscillation.\\
We will study the dynamical Casimir effect for the transverse
electric modes. Before applying our formalism, let us first write
$TE$ modes in static case. In the mode $TE$, for $\ds{t<0}$ the
cavity is static, and each mode is given by
\begin{eqnarray}
A_{x}&=&~~0\nonumber\\
A_{y}&=&\sum_{n_x,n_y,n_z}A_{0_{y}}~\exp{(-i\omega_{n_xn_yn_z}t)}\sin
\left(\frac{n_x \pi x}{L_{x}}\right)
\cos\left(\frac{n_y\pi y}{L_{y}} \right)\sin \left(\frac{n_z\pi z}{L_{z}}\right)\nonumber\\
A_{z}&=&\sum_{n_x,n_y,n_z}A_{0_{z}}~\exp{(-i\omega_{n_xn_yn_z}t)}\sin
\left(\frac{n_x\pi x}{L_{x}}\right)\sin \left(\frac{n_y\pi
y}{L_{y}}\right)\cos \left(\frac{n_z\pi z}{L_{z}}\right)~,
\end{eqnarray}
where $n_x,n_y,n_z$ are positive integers. The constants
$\ds{A_{0_{y}}}$ and $\ds{A_{0_{z}}}$ satisfy the coulomb gauge condition, $\ds{A_{0_{y}}n_y/L_{y}+A_{0_{z}}n_z/L_{z}=0}$. \\
At time $t=0$, the rectangular cavity resonator starts to
oscillate in the x-direction as a whole. The positions of the six
walls are given by (\ref{L1L23D}). Let us find the components of
the field operator at any time. We will study with the moving
coordinate systems $(q,y,z)$. The field operator $\ds{A(q,y,z,t)}$
associated with a vector potential satisfies the transformed
three-dimensional wave equation (c=1)
\begin{equation}\label{vectors6}
\frac{1}{L_x^2 }~ \frac{\partial^2 \vec{A}}{\partial
{q}^{2}}+\frac{\partial^2 \vec{A}}{\partial
{y}^{2}}+\frac{\partial^2 \vec{A}}{\partial {z}^{2}} =
\frac{\partial^2 \vec{A}}{\partial {t}^{2}}- \epsilon\left(
2\Omega  ~ \cos \Omega t~ \frac{\partial^2 \vec{ A}}{\partial {q}
\partial {t}}- \Omega^2   \sin \Omega t ~\frac{\partial
\vec{A}}{\partial {q}} \right)~.
\end{equation}
When $\ds{t>0}$, the solution of the components of the field
operator may be expanded in terms of the orthogonal basis
functions.
\begin{eqnarray}\label{gdhjki}
A_{x}&=&0\nonumber\\
A_{y}&=&\sum_{n_x,n_y,n_z}A_{0_{y}}~a_{n_xn_yn_z}^{k_{x}k_{y}k_{z}}(t)~\sin \left(n_x\pi q \right) \cos\left(\frac{n_y\pi y}{L_{y}} \right)\sin \left(\frac{n_z\pi z}{L_{z}}\right)~,\nonumber\\
A_{z}&=&\sum_{n_x,n_y,n_z}A_{0_{z}}~a_{n_xn_yn_z}^{k_{x}k_{y}k_{z}}(t)~\sin
\left(n_x\pi q\right)\sin \left(\frac{n_y\pi y}{L_{y}}\right)\cos
\left(\frac{n_z\pi z}{L_{z}}\right)~,
\end{eqnarray}
where the time dependent function
$\ds{a_{n_xn_yn_z}^{k_{x}k_{y}k_{z}}(t)}$ is to be determined
later. Let us substitute the equation (\ref{gdhjki}) into the
transformed wave equation (\ref{vectors6}). Using the
orthogonality relations, we obtain the dynamical equations.
\begin{equation}\label{denklem0hjdk0}
\ddot{a}_{m_{x}n_{y}n_{z}}^{k_{x}k_{y}k_{z}}+\omega_{m_{x}n_{y}n_{z}}^{2}
a_{m_{x}n_{y}n_{z}}^{k_{x}k_{y}k_{z}}=\epsilon\left( 4\Omega \cos
(\Omega t) \sum_{n_x\neq
m_x}g_{n_{x}m_{x}}~\dot{a}_{n_{x}n_{y}n_{z}}^{k_{x}k_{y}k_{z}}-
2\Omega^{2} \sin (\Omega t) \sum_{n_x\neq
m_x}g_{n_{x}m_{x}}~a_{n_{x}n_{y}n_{z}}^{k_{x}k_{y}k_{z}} \right)
\end{equation}
As can be seen, this equation and the equation (\ref{denklem00gf})
for the scalar field are the same. So, for the dynamical Casimir
effect problem, both scalar and vector field cases can be treated
in the similar way. Multiple scale analysis has already been
applied to the scalar field. So, the equations
(\ref{yhfbcnjk},\ref{yhfbcnjk2}) are also valid for $TE$ modes. We
will study the photon
generation for the $TE$ modes by giving an example.\\
We demand that the generated photons have the modes $\ds{(2,2,1)}$
and $\ds{(3,2,1)}$. These two modes are coupled if we choose $\ds{\Omega=(3+\sqrt{14})\pi/L_x}$.\\
After do the same things with the help of MSA we obtain,
\begin{eqnarray}
\left(%
\begin{array}{c}
  A_{2,2,1}^{{(k_{x}k_{y}k_{z})}} \\
  A_{3,2,1}^{{(k_{x}k_{y}k_{z})}} \\
\end{array}%
\right)=
\left(%
\begin{array}{c}
  \frac{1}{\sqrt{2\omega_{2,2,1}}}~\delta_{k_x,2}\cosh(\lambda^{\prime}\tau)\\
  \frac{1}{\sqrt{2\omega_{3,2,1}}}~\delta_{k_x,3}\cosh(\lambda^{\prime}\tau)\\
\end{array}%
\right)~;~~~~~~\left(%
\begin{array}{c}
  B_{2,2,1}^{{(k_{x}k_{y}k_{z})}} \\
  B_{3,2,1}^{{(k_{x}k_{y}k_{z})}} \\
\end{array}%
\right)=
\left(%
\begin{array}{c}
  \frac{G^{-}_{(3,2,1),(2,2,1)}}{\lambda^{\prime} \sqrt{2\omega_{2,1,1}}}~\delta_{k_x,2}\sinh(\lambda^{\prime}\tau)\\
  \frac{G^{-}_{(2,2,1),(3,2,1)}}{\lambda^{\prime}\sqrt{2\omega_{1,1,1}}} ~\delta_{k_x,3}\sinh(\lambda^{\prime}\tau)\\
\end{array}%
\right)~,
\end{eqnarray}
where
$\ds{{\lambda^{\prime}}^{2}=G^{-}_{321,221}~G^{-}_{221,321}}$.
Then, we calculate the number of generated photons (\ref{number1})
\begin{eqnarray}
\langle N_{2,2,1}\rangle &=& ~\sinh^{2}(\lambda^{\prime}\tau)\nonumber\\
\langle N_{3,2,1}\rangle &=& ~\sinh^{2}(\lambda^{\prime}\tau)~.
\end{eqnarray}
Here, we have performed analytical calculations to find the number
of generated photons by using MSA. To this end, it should be
mentioned that Ruser found perfect agreement between the numerical
results and analytical predictions obtained by MSA \cite{ruser}.

\section{Enhancement of Photon Numbers}

So far, we have considered the cavity resonator oscillated as a
whole. We will now study the case of symmetric oscillation with
respect to the center of the cavity in the $x$ direction.
\begin{equation}\label{L1Ldh23D}
L_{1x}(t) = -\epsilon ~L_x\sin \Omega t~,~~ L_{2x}(t) =
L_x\left(1+\epsilon~ \sin \Omega
t~\right)~,~~L_{1y}(t)=0~,~~L_{2y}(t)=L_y~,~~L_{1z}(t)=0~,~~L_{2z}(t)=L_z~.
\end{equation}
In this case, the volume of the cavity changes in time. The two
walls in the $x$ direction move opposite to each other. \\
For this configuration, only the scalar field will be treated.
This is because, as was pointed above, the equations for
$\ds{a_{{n_xn_yn_z}}^{{k_xk_yk_z}}(t)}$ are the same for the
scalar field and the vector field. In other words, the number of
produced TE-mode photons equals the number of produced scalar
particles in a three dimensional cavity. So, it is enough to study
the dynamical Casimir effect for the scalar field to understand
the underlying mechanism. For an arbitrary moment of time, the
mode function for the scalar field is expanded as
\begin{equation}
\Psi_{k_xk_yk_z}(t>0)=\sum_{n_xn_yn_z}
a_{{n_xn_yn_z}}^{{k_xk_yk_z}}(t)~\sqrt{\frac{L_x}{L_{2x}-L_{1x}}}~\sin(n_x\pi
q)~\sin(\frac{n_y\pi}{L_y} y)~\sin(\frac{n_z\pi}{L_z} z)~,
\end{equation}
where $\ds{q(t)=\frac{x-L_{1x}}{L_{2x}-L_{1x}}}$. If we substitute
it into the wave equation and use the orthogonality relations, we
get
\begin{equation}\label{ildenklem00gf}
\ddot{a}_{m_{x}n_{y}n_{z}}^{k_{x}k_{y}k_{z}}+\omega_{m_{x}n_{y}n_{z}}^{2}(t)~
a_{m_{x}n_{y}n_{z}}^{k_{x}k_{y}k_{z}}=\epsilon\left(4\Omega \cos
(\Omega t) \sum_{n_x\neq
m_x}g_{n_{x}m_{x}}~\dot{a}_{n_{x}n_{y}n_{z}}^{k_{x}k_{y}k_{z}}-2\Omega^{2}
\sin (\Omega t) \sum_{n_x\neq
m_x}g_{n_{x}m_{x}}~a_{n_{x}n_{y}n_{z}}^{k_{x}k_{y}k_{z}} \right)
\end{equation}
where~
$\ds{\omega_{m_{x}n_{y}n_{z}}^{2}(t)=\pi^2(~{m_{x}^2/(L_{2x}-L_{1x})^2+n_{y}^2/L_y^2+n_{z}^2/L_z^2}~)}$
and the new antisymmetric coefficient is given by $\ds{
g_{n_xm_x}=\frac{m_xn_x(1+(-1)^{m_x+n_x} )}{n_x^2-m_x^2}}$ for
$\ds{m_x\neq n_x}$. \\
There are two differences between (\ref{denklem00gf}) and
(\ref{ildenklem00gf}).  Firstly, the antisymmetric coefficient
vanishes when $\ds{m_x+n_x}$ is an even number when the cavity
oscillates as a whole (given below (\ref{denklem00gf})). However,
it vanishes when $\ds{m_x+n_x}$ is an odd number when the cavity
oscillates symmetrically. Secondly, the term
$\ds{\omega_{m_{x}n_{y}n_{z}}^{2}}$ in the left hand side of the
former one is constant while it is time dependent for the latter
one. The time dependent character of it gives a modification of
the equation (\ref{uhdn886})
\begin{eqnarray}\label{iluhdn886}
\ddot{a}_{m_{x}n_{y}n_{z}}^{{k_{x}k_{y}k_{z}(1)}} +
\omega_{m_{x}n_{y}n_{z}}^2~
a_{m_{x}n_{y}n_{z}}^{{k_{x}k_{y}k_{z}(1)}}=&&-2\partial_{\tau
}\partial_{
t}~a_{m_{x}n_{y}n_{z}}^{{k_{x}k_{y}k_{z}(0)}}+4\frac{\pi^2
m_x^2}{L_x^2}\sin (\Omega
t)~a_{m_{x}n_{y}n_{z}}^{{k_{x}k_{y}k_{z}(0)}}+ 4\Omega \cos
(\Omega t) \sum_{n_x\neq
m_x}g_{n_{x}m_{x}}~\dot{a}_{n_{x}n_{y}n_{z}}^{k_{x}k_{y}k_{z} (0)}
\nonumber\\&&- 2\Omega^{2} \sin (\Omega t) \sum_{n_x\neq
m_x}g_{n_{x}m_{x}}~a_{n_{x}n_{y}n_{z}}^{k_{x}k_{y}k_{z}(0)}
\end{eqnarray}
The second term in the right hand side is new. Let us study the
the parametric resonance case,
$\ds{\Omega=2\omega_{m_{x}n_{y}n_{z}}}$. Then, MSA gives equations
for $A(\tau)$ and $B(\tau)$
\begin{eqnarray}\label{ilyhfbcnjk}
\partial_{\tau}A_{m_{x}n_{y}n_{z}}^{{k_{x}k_{y}k_{z}}}+\frac{\pi^{2}m_{x}^{2}}{L_{x}^{2}~\omega_{m_x}}B_{m_{x}n_{y}n_{z}}^{{k_{x}k_{y}k_{z}}}
+
G^{-}_{(n_{x}n_{y}n_{z}),(m_{x}n_{y}n_{z})}~A_{n_{x}n_{y}n_{z}}^{{k_{x}k_{y}k_{z}}}
&=&0~~~~~~~~\\\nonumber\\
\label{9850njk2}
-\partial_{\tau}B_{m_{x}n_{y}n_{z}}^{{k_{x}k_{y}k_{z}}}-\frac{\pi^{2}m_{x}^{2}}{L_{x}^{2}~\omega_{m_x}}A_{m_{x}n_{y}n_{z}}^{{k_{x}k_{y}k_{z}}}
-G^{-}_{(n_{x}n_{y}n_{z}),(m_{x}n_{y}n_{z})}~B_{n_{x}n_{y}n_{z}}^{{k_{x}k_{y}k_{z}}}
&=&0~~~~~~~
\end{eqnarray}
where $\ds{G^{-}_{(n_{x}n_{y}n_{z}),(m_{x}n_{y}n_{z})}}$ was
defined in (\ref{scalarGmn}) and $n_x$ is a positive integer which
satisfy the following relation
\begin{equation}\label{hid6hwsw}
\omega_{n_{x}n_{y}n_{z}}=3\omega_{m_{x}n_{y}n_{z}}~.
\end{equation}
Let us give some examples. Firstly, we are interested in the
uncoupled modes. For example, consider the mode $\ds{(1,1,0)}$. If
we solve (\ref{ilyhfbcnjk},\ref{9850njk2}) endowed with the
initial conditions (\ref{sgdhjk}), we get
\begin{eqnarray}\label{ilyhfnmbcnjk}
A_{110}^{110}(\tau)&=&\frac{1}{\sqrt{2\omega_{110}}}~\cosh(\lambda \tau)\\\nonumber\\
\label{ks4jk2}
B_{110}^{110}(\tau)&=&-\frac{1}{\sqrt{2\omega_{110}}}~\sinh(\lambda
\tau)~,
\end{eqnarray}
where $\ds{\lambda=\frac{~\pi}{\sqrt{2}L_{x}}}$. The number of
photons is calculated (\ref{number1})
\begin{equation}
<N_{1,1,0}>=\sinh^{2}(\lambda\tau)~.
\end{equation}
Compare this result with the one obtained in \cite{z27} where the
authors assumed that only one wall is oscillating in the
parametric resonance case. They found that
$\ds{<N_{1,1,0}>=\sinh^{2}(\lambda_D\tau)}$, where
$\ds{\lambda_D=\lambda/2}$.\\
Number of generated photons are the same for the following two
systems: The cavity with single oscillating mirror with $\ds{2
\epsilon}$ and the cavity with two oppositely oscillating mirrors
with $\ds{ \epsilon}$. Instead of increasing $\ds{ \epsilon}$ by
factor $2$, the static wall is allowed to oscillate as described
above.\\
As a second example consider the mode $\ds{(1,1,1)}$. In this
case, according to (\ref{hid6hwsw}), the mode $\ds{(5,1,1)}$ is
coupled. The solutions of (\ref{ilyhfbcnjk},\ref{9850njk2}) become
\begin{eqnarray}\label{ilfg0njk}
A_{511}^{511}(\tau)&=&\frac{1}{\sqrt{2\omega_{511}}}\delta_{5,k_{x}}\cosh(\lambda_{1} \tau)\\\nonumber\\
\label{dfcbmxnjk2}
B_{511}^{511}(\tau)&=&-\frac{1}{\sqrt{2\omega_{511}}}\delta_{5,k_{x}}\sinh(\lambda_{1}\tau),
\end{eqnarray}
\begin{eqnarray}\label{hgyuynmjk}
A_{~111}^{k_{x}11}~(\tau)&=&0.681\frac{\delta_{k_{x},5}}{\sqrt{2\omega_{511}}}(\sinh(\lambda_{1}\tau)-\sinh(\lambda_{2}\tau))
+\frac{\delta_{k_{x},1}}{\sqrt{2\omega_{111}}}\cosh(\lambda_{2}\tau)\nonumber\\
B_{~111}^{k_{x}11}~(\tau)&=&0.681\frac{\delta_{k_{x},5}}{\sqrt{2\omega_{511}}}(\cosh
(\lambda_{2} \tau)-\cosh(\lambda_{1} \tau))-
\frac{\delta_{k_{x},1}}{\sqrt{2\omega_{111}}}\sinh(\lambda_{2}\tau)
\end{eqnarray}
where $\ds{\lambda_{1}=\frac{\pi}{L_x} \frac{25}{\sqrt{27}}}$ and
$\ds{\lambda_{2}=\frac{\pi}{L_x} \frac{1}{\sqrt{3}}}$. Then, the
number of photons
\begin{eqnarray}\label{ilyhfbcnjkazhznghj}
\langle N_{1,1,1}\rangle &=& ~\sinh^{2}(\lambda_{2}\tau)+0.1549~(\cosh^{2}(\lambda_{2}\tau)+\cosh^{2}(\lambda_{1}\tau)-2\cosh^{2}(\lambda_{2}\tau)\cosh^{2}(\lambda_{1}\tau))\nonumber\\
\langle N_{5,1,1}\rangle &=& ~\sinh^{2}(\lambda_{1}\tau)~.
\end{eqnarray}
As a result, compared to the result obtained for a single mirror,
the radiated photon flux is enhanced.

\section{Discussion}

Kim, Brownell and Onofrio proposed an experiment for the detection
of the dynamical Casimir effect \cite{qw1}. They considered a
three dimensional cavity with a single moving boundary. Taking
into account the limitation by the photon leakage of the cavity
expressed through the optical quality factor $Q_{opt}$, which
saturates at the hold time $\ds{\tau=Q_{opt}/\omega}$, they gave
formula for the maximum photon population for the parametric case
and uncoupled mode as
\begin{equation}\label{hidyhs5}
<N>=\sinh^{2}(2 ~Q_{opt}~\epsilon~ t)~,
\end{equation}
We have shown that if the two walls are moving opposite to each
other with the same frequencies and amplitudes, $\ds{\epsilon}$
should be replaced by $\ds{2\epsilon}$. Hence
\begin{equation}\label{hidyhs5g}
<N>=\sinh^{2}(4 ~Q_{opt}~\epsilon~ t)~,
\end{equation}
The number of generated photons in the cavity is very sensitive to
the product $\ds{\epsilon ~Q_{opt}}$. In current technology, the
maximum value $\ds{\epsilon ~Q_{opt}=1}$ \cite{qw1}. The equation
(\ref{hidyhs5}) gives  the number of generated photons $\ds{N=
13}$ if only one wall is in motion. If two walls move
symmetrically with respect to the center of the cavity,
(\ref{hidyhs5g}) gives the number of generated photons
$\ds{N=  745}$.\\
The difference between the two cases is great if $\ds{\epsilon
~Q_{opt}=2}$, which may be possible in the future. In this case,
the first formula (\ref{hidyhs5}) gives $\ds{N= 745}$. However,
the second one (\ref{hidyhs5g}) gives large number of
photons $\ds{N= 2.2 ~10^6}$.\\
From the experimental point of view, we think that the systems
with two moving walls will play an important role for the
detection of generated photons.

\end{document}